\documentclass[preprint,showpacs,preprintnumbers,amsmath,amssymb]{revtex4}
\usepackage{graphicx}
\usepackage{dcolumn}
\usepackage{bm}
\usepackage{xcolor}

\newcommand{\be}{\begin{equation}}
\newcommand{\en}{\end{equation}}
\newcommand{\bea}{\begin{eqnarray}}
\newcommand{\ena}{\end{eqnarray}}

\begin{document}


\title{    G-inflation: From the intermediate,  logamediate  and exponential models  }

\author{Ram\'on Herrera}
\email{ramon.herrera@pucv.cl}
\author{Nelson Videla}
\email{nelson.videla@pucv.cl} \affiliation{ Instituto de
F\'{\i}sica, Pontificia Universidad Cat\'{o}lica de Valpara\'{\i}so,
Avenida Brasil 2950, Casilla 4059, Valpara\'{\i}so, Chile.}

\author{Marco Olivares}
\email{marco.olivaresr@mail.udp.cl} \affiliation{ Facultad de
    Ingenier\'{i}a y Ciencias, Universidad Diego Portales, Avenida Ej\'{e}rcito Libertador 441, Casilla 298-V, Santiago, Chile.}

\date{\today}

\begin{abstract}

The intermediate, logamediate and exponential inflationary models
in the context of Galileon inflation or G-inflation are studied.
By assuming a coupling of the form
$G(\phi,X)\propto\phi^{\nu}\,X^{n}$ in the action, we obtain
different analytical solutions from the background cosmological
perturbations assuming the slow-roll approximation. General
conditions required for these models of G-inflation to be
realizable are determined and discussed. In general, we analyze
the condition of inflation and also we use recent astronomical and
cosmological
 observations  for constraining the parameters appearing  in these G-inflationary models.
\end{abstract}

\maketitle
\flushbottom

\newpage

\section{Introduction }
It is well known that the inflationary epoch
\cite{Starobinsky:1980te,R1,R106,R103,R104,R105,Linde:1983gd}
provides more than the mechanism for solving the problems of the
hot big bang model  (flatness, horizon etc). In this sense, one of
the achievements of the inflationary universe is to provide the
primordial curvature perturbations, which seed the observed cosmic
microwave background (CMB) temperature anisotropies
\cite{astro,astro2,astro202,Hinshaw:2012aka,Ade:2013zuv,Ade:2013uln,Ade:2015xua,Ade:2015lrj}
and the structure formation of the universe, that are generated
from vacuum fluctuations of the scalar field which drives the
accelerated expansion
\cite{Starobinsky:1979ty,R2,R202,R203,R204,R205}. One can test the
inflationary paradigm by comparing the theoretical predictions for
various models of inflation with current astrophysical and
cosmological observations, in particular those that come from the
CMB temperature anisotropies. In doing so, the predictions of
representative inflationary models, given on the $n_s-r$ plane,
are compared with the allowed contour plots from the observational
data. In this context, the BICEP2/Keck-Array collaboration
\cite{Array:2015xqh} published new more precise data regarding the
CMB temperature anisotropies, improving the upper bound on the
tensor-to-scalar ratio to be $r_{0.05} < 0.07$ ($95\%$ CL) in
comparison to latest data of Planck \cite{Ade:2015lrj}, for which
$r_{0.002} < 0.11$ ($95\%$ CL).

On the other hand, in the context of exact inflationary solutions,
one of the more interesting are found by using an exponential
potential for the inflaton, yielding a power-law evolution of the
scale factor in cosmic time, i.e., $a(t)\propto t^p$, where $p>1$
\cite{Lucchin:1984yf}. Another exact solution corresponds to de
Sitter inflation in which the effective potential is a  constant
\cite{R1}. We also have an exact solution for an inverse power-law
potential. Here,  the inflationary stage can be described by the
intermediate inflation model, in which the scale factor has the
following dependence on cosmic time
\cite{Barrow:1990vx,Barrow:1993zq,Barrow:2006dh}.
\begin{equation}
a(t)=\exp\left[A\,t^f\right],\label{aa1}
\end{equation}
where $A$ and $f$ are constant parameters, satisfying the
conditions $A>0$ and $0<f<1$. This intermediate expansion law
becomes slower than de Sitter inflation, but faster than power-law
inflation instead. In addition, a generalized inflation model is
provided by the model of  logamediate inflation, in which the
scale factor evolves as \cite{Barrow:2007zr}
\begin{equation}
a(t)=\exp\left[B\,\ln(t)^{\lambda}\right],\label{loga}
\end{equation}
here, $B$ and $\lambda$ are dimensionless constant parameters such
that $B>0$ and $\lambda>1$. Note that for the special case
$\lambda=1$ and $B=p$, the logamediate inflation model reduces to
power-law inflation with  an exponential potential
\cite{Lucchin:1984yf}.

Originally, these inflationary models were studied as exact
solutions of background evolution. However, the slow-roll
formalism provides a better analysis regarding the dynamics of
primordial perturbations. In practice, these models are completely
ruled out by current observational data \cite{Ade:2015lrj} in the
standard canonical inflationary scenario. In particular, for the
intermediate inflation model, it was found that for the special
case $f=2/3$, the scalar spectral  index becomes $n_s= 1$,
corresponding to the Harrison-Zel'dovich spectrum, being not
supported by current data. Also, an  observational consequence is
that for both inflationary models,  the tensor-to-scalar ratio
$r$, becomes significantly $r\neq 0$, but  this ratio is always
$r>0.1$, as it was shown in \cite{Barrow:2007zr, Barrow:2006dh}.
If we go further the standard cold scenario, e.g., in the warm
inflation scenario, both intermediate and logamediate models may
be reconciled with current observations available at that time
\cite{Herrera:2016sov,Herrera:2015aja,Herrera:2014mca,Herrera:2013rra}.

Instead of considering the  parametrization of the scale factor as
function the cosmic time, alternatively the authors in
Ref.\cite{Myrzakulov:2014pfa} introduced an explicit expression
for the Hubble rate. Here, they studied a Hubble parameter having
an exponential dependence on cosmic time of the form
\begin{equation}
H(t)=\alpha \exp[-\beta t],\label{Hexp}
\end{equation}
where $\alpha$ denotes the value of the Hubble rate when cosmic
time tends to zero and $\beta$ is a constant parameter, such that
$\beta>0$. On the contrary of the intermediate and logamediate
inflation models, this exponential Hubble rate has the novelty of
addressing the end of inflation\cite{Myrzakulov:2014pfa}.
Nevertheless, regarding the predictions for this model on the
$n_s-r$ plane, the trajectory lie outside the 95 $\%$ CL region,
being completely ruled out by current observations.

On the other hand, going  beyond the standard canonical inflation
scenario, a non-canonical inflation model, whose Lagrangian
contains higher derivative terms,  has become of a special
interest from the theoretical and observational points of view,
yielding a large or small amount of non-Gaussianities and a
non-trivial speed of sound. A special class of such a models,
dubbed Galileon inflation models or G-inflation, were inspired by
theories exhibiting ``Galilean'' symmetry,
$\partial_{\mu}\phi\,\rightarrow\,\partial_{\mu}\phi+b_{\mu}$\cite{Nicolis:2008in}.
Interestingly, the field equations derived from such a theories
still contain derivatives up to second order, avoiding ghosts
\cite{Nicolis:2008in}. Nevertheless, this feature holds only when
the space-time is Minkowsi \cite{Deffayet:2009wt}. Although the
``covariantization'' of the Galileon achieved the equations of
motion to keep of second order, the Galilean invariance is broken
\cite{Deffayet:2011gz,Deffayet:2009wt}. This theory, as it was
shown in \cite{Charmousis:2011bf} and \cite{Kobayashi:2011nu}, is
equivalent to Horndeski's theory \cite{Horndeski:1974wa}, which is
stated as the most general scalar-tensor theory with second-order
field equations. For a representative list of works on
G-inflation, see
Refs.\cite{Kobayashi:2010cm,Burrage:2010cu,Gao:2011qe,Kobayashi:2011pc,Kamada:2010qe,Ohashi:2012wf,Unnikrishnan:2013rka,Hirano:2016gmv,BazrafshanMoghaddam:2016tdk,Herrera:2017qux,Herrera:2018mvo}.

In the framework of modified gravity theories having extra degrees
of freedom, the action for linearized gravitational waves (GWs)
reads $ S_h=\frac{1}{2}\int\,d^3x dt
M^2_{*}\left[\dot{h}^2_A-c^2_T\left(\nabla h_A\right)^2\right],$
where $M_*$ is an effective Planck mass which would depend on the
particular theory under consideration, and $h_A$ are the
amplitudes of the polarization states of the perturbations $h_{\mu
\nu }$ around the Minkowski space. The quantity $c_T$ corresponds
to the speed of the GW, which can be parameterized more convenient
as $c^2_T=1+\alpha_T$. By combing the gravitational wave event
GW170817 \cite{TheLIGOScientific:2017qsa}, observed by the
LIGO/Virgo collaboration, and the gamma-ray burst GRB 170817A
\cite{Monitor:2017mdv}, it has been possible to strongly constrain
the speed of GWs, determining that GWs propagate at the speed of
light with $\left|\alpha_T\right|\lesssim 10^{-15}$
\cite{Baker:2017hug}. However, we mentioned that this constraint
on the speed of GWs occurs
 for a redshift $z\sim 0.1$, wherewith this constraint
 does not necessarily apply to the early universe.
 As a direct consequence for Horndeski's
theory, is that a large model space of this theory has been
eliminated to the present time. Specifically, all the terms that
lead to non-minimal
 kinetic couplings are ruled out, leaving this theory constructed only with k-essence, cubic Galileon and
non-minimally coupling sectors, in which the Lagrangian density
can be written as \cite{Baker:2017hug,Langlois:2017dyl}
\begin{equation}
\label{LG} \mathcal{L}=K(\phi,X)-G(\phi,X)\Box \phi+f(\phi)R.
\end{equation}

In \cite{Teimoori:2017kob}, the authors
explored the viability of considering the intermediate inflation
model in the framework of G-inflation, with a
 cubic Galileon term of the form $G(\phi,X)\Box \phi \propto X^n\Box \phi$. Interestingly,
 it was found the compatibility of this model with Planck 2015
 data. Here the authors find that the power $n$ plays a fundamental role on   the cosmological parameters in order to obtain the observational data.




The main goal of the present article is to explore the
observational consequences of studying the intermediate,
logamediate and exponential Hubble inflation models in the
framework of the cubic Galileon and how these models are modified
with the coupling $G(\phi,X)$. In doing so, we consider a coupling
of the form $G(\phi,X)\propto \phi^{\nu}X^n$, which generalizes
the cases $G(\phi,X)\propto \phi X$ and $G(\phi,X)\propto X^n$
already studied
 in Refs.\cite{Ohashi:2012wf} and \cite{Teimoori:2017kob}, respectively.
  We will show that, for each inflation model studied, there exist a region in the space of parameters for
   which its predictions lie inside the allowed region from BICEP2/Keck-Array data, resurrecting these inflationary models. In addition, we will show
    that the allowed region in the space of parameters becomes different
     than the obtained in the case of intermediate model \cite{Teimoori:2017kob}.
 Here,  following Ref.\cite{MR} the authors of
 \cite{Teimoori:2017kob},
 introduce an extra time that corresponds to a time of an unspecified
 reheating mechanism in order to induce to stop inflation and so
 evaluate the cosmological parameters.

We have organized this article as follows. In the next section, we
present a brief review of G-inflation. In sections \ref{Gint},
\ref{Glog}, and \ref{Gexp} we study the background and
perturbative dynamics of our concrete inflationary models under
the slow-roll approximation. Contact between the predictions of
the model and observations will be done by computing the power
spectrum, the scalar spectral index as well as the
tensor-to-scalar ratio. We summarize our findings and present our
conclusions in Section \ref{conclu}. We chose units so that
$c=\hbar=8\pi G=1$.

\section{G-INFLATION}
In this section we give a brief review on the background dynamics and the cosmological perturbations
 in the model of  G-inflation. Our starting point, is
 the 4-dimensional action  in the framework of the Galilean model  given by
 \be
S=\int\sqrt{-g_4}\left( {R \over 2}+K(\phi, X)-G(\phi, X)\,\Box \phi\right) d^4x
\,.\label{accion}
 \en
 Here, the quantity   $g_{4}$ corresponds to  the determinant of the space-time
metric $g_{\mu\nu}$,  $R$ denotes  the Ricci
scalar and   $X=-g^{\mu\nu}\partial_{\mu}\phi\partial_{\nu}\phi/2$. The scalar field
is denoted by  $\phi$  and
the quantities   $K$ and $G$ are arbitrary functions of $X$ and
$\phi$.

By assuming a spatially flat Friedmann Robertson Walker (FRW) metric
 and a homogeneous scalar field $\phi=\phi(t)$,  then
 the modified Friedmann equations can be written as
\be
3H^2+K+\dot{\phi}^2(G_{\phi}-K_{X})-3HG_{X}\,\dot{\phi}^3=0,
\,\label{frwa}
 \en
 and
 \be
 2\dot{H}+3H^2+K-\dot{\phi}^2(G_{\phi}+G_{X}\ddot{\phi})=0,
 \,\label{frwb}
 \en
 where $H=\frac{\dot{a}}{a}$ corresponds to Hubble rate and $a$ denotes the
 scale factor. In the following,  we will consider that  the dots denote
differentiation with  respect to cosmic time and the notation
$K_X$ denotes
 $K_X=\partial K/\partial X$, while $K_{XX}$ corresponds to
 $K_{XX}=\partial ^2K/\partial X^2$, and $G_\phi$ means $G_\phi=\partial G/\partial\phi$, etc.

From variation of the action (\ref{accion})  with respect to
the scalar field we have
$$
3\dot{H}G_{X}\,\dot{\phi}^2+\ddot{\phi}\left[3HG_{XX}\,\dot{\phi}^3
-\dot{\phi}^2(G_{\phi X}-K_{XX})+6HG_{X}\,\dot{\phi}-2G_{\phi}+K_{X}\right]+
$$
\be 3HG_{\phi X}\,\dot{\phi}^3+\dot{\phi}^2(9H^2G_{X}-G_{\phi \phi
}+K_{\phi X})-K_{\phi} -3H\dot{\phi}(2G_{\phi }-K_{X})=0.
\,\label{frwc} \en In  the specific  cases  in which the functions
$K=X-V(\phi)$ (with $V(\phi)$ being the effective potential for
the scalar field) and $G=0$, General Relativity (GR) is recovered.

In order to study the model of   G-inflation from different inflationary expansions,
 we will analyze  the specific  case  in which  the functions $K(\phi,X)$ and $G(\phi,X)$
are given by
 \be
K(\phi, X)=X-V(\phi), \qquad \text{and}  \qquad G(\phi, X)=g(\phi)\,X^n,
 \,\label{kyg}
 \en
 respectively. Here, the coupling $g(\phi)$ is a function that depends exclusively on the scalar
field $\phi$ and the power $n$  is such that $ n>0$. Also, in the
following we will assume a power-law dependence on the scalar
field for the coupling \be g(\phi)=\gamma\,\phi^{\nu}
\,,\label{gfi} \en where the parameter $\gamma$ and the power
$\nu$ are  both real, with $\gamma>0$. Thus, the function
$G(\phi,X)$ is defined as $G(\phi, X)=\gamma\,\phi^{\nu}\,X^n$ and
then the Galilean term in the action is
$G(\phi,X)\square\phi\propto\phi^\nu\,X^{n}\,\square\phi$. We
mention that for the particular case in which $\nu=0$ i.e.,
$g(\phi)=$const., and therefore the function $G(\phi,X)\propto
X^n$ was already analyzed in Ref.\cite{Teimoori:2017kob} for the
specific model of intermediate inflation.

Following Ref.\cite{Kobayashi:2011pc}, we will consider the model
of G-inflation under the slow-roll approximation. In this sense,
the effective potential dominates over the functions $X$,
$|G_{X}H\dot{\phi} ^3|$ and $|G_{\phi}X|$. Thus,  under this
approach, the Friedmann equation
 given by Eq.(\ref{frwa}) can be approximated to
  \be
 3H^2\approx V(\phi).
 \,\label{frwa2}
 \en
 By assuming the  slow-roll approximation, we can
 introduce  the set of slow-roll parameters for G-inflation, defined as \cite{Kobayashi:2011pc}
$$
\delta_X={K_X X\over H^2},
\quad \delta_{GX}={G_{X} \dot{\phi}X\over H},\;\;\;\;\;
\quad \delta_{G\phi}={G_{\phi} X\over H^2},
$$
\be
\varepsilon_1=-{\dot{H}\over H^2},
\quad \epsilon_2=-{\ddot{\phi}\over H\dot{\phi}}=-\delta_{\phi},
\quad \epsilon_3={g_{\phi}\dot{\phi}\over g H},
 \quad \text{and}
 \quad \epsilon_4={g_{\phi \phi}X^{n+1}\over V_{\phi}}.
\,\label{srp} \en
From the  parameters defined above and combining
 with the Friedmann equations (\ref{frwa}) and
(\ref{frwb}), the slow-roll parameter $\varepsilon_1$ can be
rewritten as \be
 \varepsilon_1=\delta_X+3 \delta_{GX}-2\delta_{G\phi}-\delta_{\phi} \delta_{G
 X}.
\,\label{epsilon1} \en Now,  from the functions $K(\phi,X)$ and
$G(\phi,X)$ given by Eq.(\ref{kyg})  and considering the slow-roll
parameters from Eqs.(\ref{srp}) and (\ref{epsilon1}), the equation
of motion for the scalar field read as
$$
3H\dot{\phi}(1-\epsilon_2/3)+
3ngX^{n-1}H^2\dot{\phi}^2(3-\varepsilon_1-2n\epsilon_2)
$$
\be
+3gX^{n-1}H^2\dot{\phi}^2[(n-1)\epsilon_3+(n+1)\epsilon_2\epsilon_3/3]
=-V_{\phi}(1-2\epsilon_4). \,\label{frwc2a} \en In the context of
the slow-roll analysis,  we are going to consider that the
slow-roll parameters $|\varepsilon_1|, |\epsilon_2|, |\epsilon_3|,
|\epsilon_4|\ll 1$, see Ref.\cite{Kobayashi:2011pc}.  In
addition, we can define other three   slow-roll
 parameters that are of second order in $\varepsilon_1$ and these are given by
 $\delta_{G\phi X}=G_{,\phi X}X^2/H^2$, $\delta_{G\phi \phi}=G_{,\phi \phi}\dot{\phi}X/H^3$, and
 $\delta_{G \phi XX}=G_{,\phi XX}X^3/H^2$, respectively. Then, the
slow-roll equation of motion for the scalar field, given by
Eq.(\ref{frwc2a}), can be approximated to

 \be 3H\dot{\phi}(1+\mathcal{A})
\simeq-V_{\phi}\,, \,\label{frwc2} \en where $\mathcal{A}$ is a
function defined as \begin{equation} \mathcal{A}\equiv
3H\dot{\phi}G_{X}=3n\,g(\phi)X^{n-1}H\dot{\phi}=3n\,\gamma\,\phi^{\nu}X^{n-1}H\dot{\phi}.\label{Ageneral}
\end{equation}

From the slow-roll equation (\ref{frwc2}),
  we may distinguish two  opposite limits. First, we have the limit $|{\cal{A}}|\ll 1$,  which corresponds to the standard  slow-roll equation in GR for the scalar field.
  However, when $ |{\cal{A}}|\gg 1$,  the Galileon term
modifies the equation for  the scalar field, and hence its
dynamics. In this context, we are interested in the latter limit
in which  the Galileon effect changes  the field dynamics. Then,
by combining Eqs.(\ref{frwa2}) and (\ref{frwc2}),  we find that
the scalar field can be written as

\begin{equation} \phi^\nu \, \dot{\phi}^{2n+1}={2^n(-\dot{H}) \over
3n\gamma\,H},\,\,\,\,\, \Rightarrow\,\,\,\,\,\,{ 2n+1
\over2n+1+\nu}\,\phi^{2n+1+\nu \over2n+1}=\left({2^n \over
3n\gamma} \right) ^{1\over2n+1} \int \left({-\dot{H} \over H}
\right) ^{1\over2n+1}dt.\label{A1}
\end{equation}
Note that this expression for $\phi(t)$ could be expressed
explicitly in terms of the  cosmic time $t$ for any model and, in
particular, for any scale factor $a(t)$ or Hubble rate $H(t)$.

 From Eq.(\ref{A1}) we obtain that the function $\cal{A}$
can be rewritten as
\begin{equation}
\mathcal{A}={3\,n\,\gamma\over 2^{\,n-1}} H\left[
{2^{\,n}(-\dot{H})\over 3\,n\,\gamma\,H}\right] ^{2n-1 \over 2n+1}
\phi^{2\nu \over 2n+1}\gg 1.
\end{equation}
Here, we have used the Friedmann equation given by(\ref{frwa2}).


On the other hand, the analysis of the cosmological perturbations
in G-inflation was developed in
Refs.\cite{Kobayashi:2010cm,Kobayashi:2011pc}. In the following,
we briefly review the basic relations governing the dynamics of
cosmological perturbations in the framework of G-inflation. In
this context, the power spectrum of the primordial scalar
perturbation $\mathcal{P}_{\mathcal{S}}$ in the slow-roll
 approximation can be written as
\cite{Kobayashi:2010cm,Kobayashi:2011pc}
 \be
\mathcal{P}_{\mathcal{S}}={H^2q_s^{1/2}\over8\pi^2 \varepsilon_s^{3/2}} \,,\label{PR}
\en where the quantities $q_s$ and $\varepsilon_s$ are defined as
\be q_s=\delta_X+2\delta_{XX}+6 \delta_{GX}+6 \delta_{GXX}-2
\delta_{G\phi} \,,\label{qs} \en and \be \varepsilon_s=\delta_X+4
\delta_{GX}-2 \delta_{G\phi}
\,,\,\,\,\,\mbox{where}\,\,\,\,\delta_{XX}={K_{XX} X^2\over H^2},
\quad\mbox{and}\,\,\,\, \delta_{GXX}={G_{XX} \dot{\phi}X^2\over
H}.\label{ves} \en
 Here, we mention that the scalar propagation speed squared  is given by
$ c_s^2={\varepsilon_s\over q_s}$.
In this form, assuming the functions given by Eq.(\ref{kyg}) and
using the slow-roll parameter $\epsilon_3$, we find that the
parameters $q_s$ and $\varepsilon_s$  are rewritten as
\begin{equation}
\label{qses} q_s={X\over H^2}\left[ 1+2n\mathcal{A}\left(1-
{\epsilon_3\over 6n^2}\right) \right],
\,\,\,\mbox{and}\,\,\,\,\,\,\varepsilon_s={X\over H^2}\left[
1+{4\over 3}\mathcal{A}\left(1- {\epsilon_3\over 4n}\right)
\right].
\end{equation}
From Eq.(\ref{PR}) and considering the above parameters, the
scalar power spectrum in the slow-roll approximation results
\cite{Kobayashi:2010cm,Kobayashi:2011pc} \be
\mathcal{P}_{\mathcal{S}}\simeq{H^4(1+2n\mathcal{A})^{1/2}\over8\pi^2
X(1+4\mathcal{A}/3)^{3/2}}
\simeq{V^3(1+\mathcal{A})^{2}(1+2n\mathcal{A})^{1/2}\over 12\pi^2
    V_{\phi}^2(1+4\mathcal{A}/3)^{3/2}}\,,\label{PR2} \en and  the scalar
propagation speed squared becomes  $ c_s^2={1+4\mathcal{A}/3\over
1+2n\mathcal{A}} \leq1$, where the power $n$ is such that
$n\geq2/3$. In the limit ${\cal{A}}\gg1$, the scalar power
spectrum, given by Eq.(\ref{PR2}), becomes approximately
\begin{equation}
\mathcal{P}_{\mathcal{S}}\simeq{3H^4\sqrt{6n}\over64\pi^2
X\mathcal{A}}\simeq{\sqrt{6n}\,V^3 \mathcal{A}\over 32\pi^2
V_{\phi}^2}.\label{P1}
\end{equation}

Also, the scalar spectral index $n_S$ associated with the tilt of
the power spectrum, is defined as  $n_s-1=d\ln{\cal{P_S}}/d\ln k$.
Thus, from Eq. (\ref{PR2}), the scalar spectral index under the
slow-roll approximation can be written as
\cite{Kobayashi:2010cm,Kobayashi:2011pc}
\begin{equation}
n_s\simeq\,1-\frac{6\epsilon}{1+{\cal{A}}}+\frac{2\eta}{1+\cal{A}}
+{\dot{\cal{A}}\over
H}\left[\frac{2}{1+{\cal{A}}}+\frac{n}{1+2n{\cal{A}}}
-\frac{2}{1+4{\cal{A}}/3}\right],\label{ns0}
\end{equation}
where $\epsilon$ and $\eta$ are the standard slow-roll parameters,
defined as
\begin{equation}
  \epsilon=\frac{1}{2}\left(\frac{V_{\phi}}{V}\right)^2,\;\,\,
\,\,\,\mbox{and}\,\,\,\,\,\,\eta=\frac{V_{\phi\phi}}{V},
\end{equation}
respectively. Here, we observe that in the limit
${\cal{A}}\rightarrow 0$ (or equivalently $g\rightarrow 0$), the
scalar spectral index given by Eq.(\ref{ns0}) coincides with the
 expression obtained in GR, where  $n_s-1\simeq -6\epsilon+2\eta$. In the
limit $|{\cal{A}}|\gg 1$, where the Galileon term dominates the
inflaton dynamics, the  scalar index $n_S$ results
\begin{equation}
n_s\simeq 1-\frac{6\epsilon}{{\cal{A}}}+ \frac{2\eta}{\cal{A}}
+{\dot{\cal{A}}\over H\cal{A}}\,\,\,.\label{ns}
\end{equation}

 On the other hand, the  tensor power spectrum in the framework of
 G-inflation is similar to  standard inflation in GR,  where the amplitude of GWs have a tensor spectrum ${\cal{P}}_G$ given by\cite{Kobayashi:2010cm,Kobayashi:2011pc}
${\cal{P}}_G={2H^2\over \pi^2 }$. In this sense, the tensor-to-scalar ratio, defined as $r={\cal{P}}_G/\mathcal{P}_{\mathcal{S}}$, in the
framework of G-inflation under slow-roll approximation can be
written as
\begin{equation}
r=\frac{{\cal{P}}_G}{{\cal{P}}_{\cal{S}}}\simeq\,16\epsilon\,\left[\frac{(1+4{\cal{A}}/3)^{3/2}}
{(1+{\cal{A}})^2(1+2n{\cal{A}})^{1/2}}\right].\label{rl}
\end{equation}
Again, we note that in the limit ${\cal{A}}\rightarrow 0$, the
tensor-to-scalar  ratio  coincides with the expression obtained in
standard inflation, where $r\simeq16\epsilon$. Now, by assuming
the limit $ |{\cal{A}}|\gg 1$, the tensor-to-scalar ratio $r$ is
approximated to
\begin{equation}
r\simeq\frac{4\sqrt{2}}{3^{3/2}}\,\frac{16\epsilon}{\sqrt{n}\,\,{\cal{A}}}\,.\label{r}
\end{equation}
Thus, at least in principle, Galileon inflation becomes
phenomenologically distinguishable from standard inflation, where
$c^2_s=1$. On the other hand, an eventual detection of
non-Gaussianities (NG), roughly measured by the non-linear
parameter $f_{NL}$, could break the degeneracy among the several
inflation models and also enables to us to  discriminate between
single-field inflation and other alternative scenarios (for a
comprehensive review see,
Refs.\cite{Bartolo:2004if,Renaux-Petel:2015bja}). In particular,
for the simplest model of inflation, consisting in a single-field
with a canonical kinetic term and a smooth inflaton potential, the
predicted amount of NG is such that $f_{NL}\ll 1$
\cite{Gangui:1993tt,Acquaviva:2002ud,Maldacena:2002vr}. Going
further the previous properties may result in a large amount of
NG, $|f_{NL}|\gg 1$, and current observational results
$f_{NL}\lesssim {\mathcal{O}}(10)$ \cite{Ade:2015ava}.

Regarding the shapes of NG, it can be determined several types
which depend on the magnitudes of the wave vectors $k_1$, $k_2$,
and $k_3$, in the Fourier space with the constraint
$k_1+k_2+k_3=0$ \cite{Babich:2004gb}. For example, multi-field
inflation \cite{Seery:2005gb} and curvaton scenarios
\cite{Sasaki:2006kq} give rise a bispectrum that has a maximum in
squeezed configuration or local shape (i.e. for $k_3\ll k_1\simeq
k_2$) \cite{Wang:1999vf,Verde:1999ij}. In particular for
non-canonical kinetic terms, the NG are well described by the
equilateral (i.e. $k_1=k_2=k_3$) and orthogonal shapes (i.e.
$k_1=2k_2=2k_3$) \cite{Chen:2006nt,Senatore:2009gt}.
An important linear combination of the equilateral and orthogonal
shapes give rise to the so-called enfolded shape and this
combination  was determined from Planck data in
Ref.\cite{Ade:2015ava}.

Following Ref.\cite{DeFelice:2013ar}, the several expressions for
non-linear parameter $f_{NL}$ have been calculated  for  the
local, equilateral, orthogonal, and enfolded configurations in the
Horndeski's most general scalar tensor theories become
\begin{eqnarray}
f_{NL}^{\textup{local}}&=&\frac{5}{12}(1-n_{s}),\label{flocal}\\
f_{NL}^{\textup{equil}}&=&\frac{85}{324}\left(1-\frac{1}{c^2_s}\right)-\frac{10}{81}\frac{\mu}{\Sigma}+
\frac{20}{81\,\varepsilon_s}(\delta_{GX}+\delta_{GXX})+\frac{65}{162\,c^2_s\,\varepsilon_s}\delta_{GX},\label{fequil}\\
f_{NL}^{\textup{ortho}}&=&\frac{259}{1296}\left(1-\frac{1}{c^2_s}\right)+\frac{1}{648}\frac{\mu}{\Sigma}-\frac{1}{324\,
\varepsilon_s}(\delta_{GX}+\delta_{GXX})+\frac{65}{162\,c^2_s\,\varepsilon_s}\delta_{GX},\label{fortho}\\
f_{NL}^{\textup{enfold}}&=&\frac{1}{32}\left(1-\frac{1}{c^2_s}\right)-\frac{1}{16}\frac{\mu}{\Sigma}+\frac{1}{8\,
\varepsilon_s}(\delta_{GX}+\delta_{GXX}),\label{fenfold}
\end{eqnarray}
respectively. Here, the expressions for $\mu$ and $\Sigma$ are
given by Eqs.(3.11) and (3.12) in reference \cite{DeFelice:2013ar}
by setting $P(\phi,X)=K(\phi,X)$, $G_{3}(\phi,X)=G(\phi,X)$, and
$G_4=G_5=0$. From Eqs.(\ref{fequil}) and (\ref{fortho}), the
enfold shape (\ref{fenfold}) is obtained as follows
\begin{equation}
\label{enfold2}
f_{NL}^{\textup{enfold}}=\frac{1}{2}(f_{NL}^{\textup{equil}}-f_{NL}^{\textup{ortho}}).
\end{equation}
Regarding the current observational constraints on primordial NG,
by combining temperature and polarization data, Planck
collaboration has found that \cite{Ade:2015ava}
\begin{eqnarray}
f_{NL}^{\textup{local}}&=&0.8\pm 5.0\,\,\,\,\,\,(68\%\,\textup{CL}),\\
f_{NL}^{\textup{equil}}&=&-4\pm 43\,\,\,\,\,\,\,\,(68\%\,\textup{CL}),\\
f_{NL}^{\textup{ortho}}&=&-26\pm 21\,\,\,\,\,(68\%\,\textup{CL}),
\end{eqnarray}
and by using Eq.(\ref{enfold2}), the current observational
constraint on $f_{NL}^{\textup{enfold}}$ becomes
\begin{equation}
f_{NL}^{\textup{enfold}}=11\pm 32\,\,\,\,\,(68\%\,\textup{CL}).
\end{equation}

For our model, in which the functions $K(\phi,X)$ and $G(\phi,X)$
are specified by Eq.(\ref{kyg}), we have that
$K_{,XX}=K_{,XXX}=0$, $G_{,\phi X}\neq 0$ and $G_{,\phi XX}\neq
0$. Then, by considering that $\delta_{G\phi X}$ and
$\delta_{G\phi XX}$ are of second order in $\epsilon_1$, the
expression for $\mu$ (Eq.(3.12) in Ref.\cite{DeFelice:2013ar})
reduces to
\begin{equation}
\mu=H^2(\delta_{GX}+5\delta_{GXX}+2\delta_{GXXX}),\label{mu model}
\end{equation}
where
\begin{equation}
\delta_{GXXX}\equiv \frac{G_{,XXX}\dot{\phi}X^3}{H}.\label{Gxxx}
\end{equation}
Note that our expression obtained for $\mu$ coincides  with those
obtained in Ref.\cite{Teimoori:2017kob}. Now, by considering that
$\delta_{XX}=0$ and $\delta_{G\phi}\neq 0$, the expression for
$\Sigma$ (Eq. (3.11) in \cite{DeFelice:2013ar}) becomes
\begin{equation}
\Sigma=H^2(\delta_x+6\delta_{GX}+6\delta_{GXX}-2\delta_{G\phi}).\label{Sigmamodel}
\end{equation}

Considering  that $G(\phi,X)=g(\phi)X^n$, from the  third slow-roll
parameter in Eq.(\ref{ves}) and Eq.(\ref{Gxxx}), we obtain the
relations $\delta_{GXX}=(n-1)\delta_{GX}$ and
$\delta_{GXXX}=(n-1)(n-2)\delta_{GX}$. In addition, by using
Eqs.(\ref{kyg}), (\ref{srp}), (\ref{Ageneral}), (\ref{qses}) and
the expressions (\ref{mu model}) and (\ref{Sigmamodel}), the
non-linear parameters for the equilateral, orthogonal, and
enfolded configuration for our particular Galileon model reduce to
\begin{eqnarray}
f_{NL}^{\textup{equil}}&=&\frac{85}{162}\frac{(2-3n){\mathcal{A}}}{3+4{\mathcal{A}}}+
\frac{{\mathcal{A}}}{243}\bigg[\frac{10n(1-2n)}{1+2n{\mathcal{A}}\left(1-\frac{\epsilon_3}{6n^2}\right)}+
\frac{60n}{3+4{\mathcal{A}}\left(1-\frac{\epsilon_3}{6n^2}\right)}\nonumber\\
&&+\frac{585(1+2n{\mathcal{A}})}{2(3+4{\mathcal{A}})\left(3+4{\mathcal{A}\left(1-\frac{\epsilon_3}{6n^2}\right)}\right)}\bigg],\label{fequilA}\\
f_{NL}^{\textup{ortho}}&=&\frac{259}{648}\frac{(2-3n){\mathcal{A}}}{3+4{\mathcal{A}}}-
\frac{{\mathcal{A}}}{486}\bigg[\frac{n(1-2n)}{4\left(1+2n{\mathcal{A}}\left(1-\frac{\epsilon_3}{6n^2}\right)\right)}+
\frac{3n}{2\left(3+4{\mathcal{A}}\left(1-\frac{\epsilon_3}{6n^2}\right)\right)}\nonumber\\
&&-\frac{585(1+2n{\mathcal{A}})}{(3+4{\mathcal{A}})\left(3+4{\mathcal{A}\left(1-\frac{\epsilon_3}{6n^2}\right)}\right)}\bigg],\label{fortholA}\\
f_{NL}^{\textup{enfold}}&=&\frac{1}{16}\frac{(2-3n){\mathcal{A}}}{3+4{\mathcal{A}}}+\frac{{\mathcal{A}}}{24}
\bigg[\frac{n(1-2n)}{2\left(1+2n{\mathcal{A}}\left(1-\frac{\epsilon_3}{6n^2}\right)\right)}+\frac{3n}{3+4{\mathcal{A}}\left(1-\frac{\epsilon_3}{6n^2}\right)}\bigg],\label{fenfoldA}
\end{eqnarray}
where  the slow-roll parameter $\epsilon_3$ is given by
$
\epsilon_3=\frac{\nu}{\phi}\left(\frac{2\varepsilon_1}{{\mathcal{A}}}\right)^{1/2},
$
since that $g(\phi)=\gamma \phi^{\nu}$ (see Eq.(\ref{gfi})).
 Note that for the particular case in
which $\epsilon_3=0$, i.e., $G(\phi,X)$ does not depend on $\phi$,
and  Eqs.(\ref{fequilA})-(\ref{fenfoldA}) reduce to those obtained
in \cite{Teimoori:2017kob}.

Assuming that the parameter $n>1$ and the slow-roll parameter
$\epsilon_3$ during inflation becomes  $\epsilon_3\ll 1$, then the
ratio $\frac{\epsilon_3}{6n^2}\ll 1$. Thus, during  the Galileon
dominated regime in which ${\mathcal{A}}\gg 1$,
the NG parameters (\ref{fequilA})-(\ref{fenfoldA}) reduce to
\begin{eqnarray}
f_{NL}^{\textup{equil}}&=&\left(\frac{275}{972}\right)-\left(\frac{865}{3888}\right)n,\label{flimit1}\\
f_{NL}^{\textup{ortho}}&=&\left(\frac{97}{486}\right)-\left(\frac{1163}{7776}\right)n,\label{flimit2}\\
f_{NL}^{\textup{enfold}}&=&\left(\frac{1}{24}\right)-\left(\frac{7}{192}\right)n.\label{flimit3}
\end{eqnarray}
Here we observe that these expressions  take the same form as
those obtained  in Ref.\cite{Teimoori:2017kob}.

Also, we find that the square of the speed of sound in this regime
becomes $c^2_s=\frac{2}{3n}$. Note that this speed only depends on
the power $n$ in the  Galileon dominated regime. In this way, for
values of $n>1$
 the speed of sound is reduced to   $c^2_s<1$, yielding
values for NG such that $ |f_{NL}| \gtrsim 1$, as it can be seen
from  Eqs.(\ref{flimit1})-(\ref{flimit3}).








In the following, we will study three different inflationary
expansions; the intermediate, logamediate and exponential  in the
framework
 of
G-inflation. In order to study  these expansions we will assume
 the Galilean effect predominates over the standard inflation, i.e., in the limit $ |{\cal{A}}|\gg 1$.

\section{Intermediate G-inflation.}\label{Gint}
Let us consider a scale factor that evolves according to
Eq.(\ref{aa1}) or commonly called intermediate expansion. Here,
the Hubble rate is given by $H(t)={Af\over t^{1-f}}$, and from
Eq.(\ref{A1}), we find that the  scalar field as a function of
cosmic time becomes
\be \phi(t)= \left({2n+1 +\nu \over 2n}
\right) ^{2n+1\over2n+1+\nu } \left({2^n \left(1-f \right)\over
3n\gamma} \right) ^{1\over 2n+1+\nu} \,t^{2n\over 2n+1+\nu}+C_0
\,,\label{sf1} \en
 where $C_0$ corresponds to an integration constant, that
without loss of generality we can take
$C_0=0$. Thus, the Hubble rate as function of the scalar
field $\phi$ becomes
$$
H(\phi)={Af\over k_1^{1-f}}\, \phi^{\,-(1-f)\mu_1},
$$
where the constants $k_1$ and $\mu_1$ are defined as
$$
k_1=\left({2n \over 2n+1 +\nu } \right) ^{2n+1\over2n}
\left({3n\gamma\over 2^n \left(1-f \right)} \right) ^{1\over 2n},
\,\,\,\, \mbox{and}\,\,\,\,  \mu_1={2n+1+\nu \over 2n},
$$
respectively. From the Friedmann equation (\ref{frwa2}), the
effective potential in terms of the scalar field can be written as
\begin{equation}
V(\phi)=3\left[{A^2 f^2\over k_1^{2(1-f)}}\right]\,
\phi^{\,-2(1-f)\mu_1},
\end{equation}
which has an inverse power-law dependence on the scalar field,
hence does not have a minimum.

 In the cosmological context, the effective potential
characterizing the canonical variables of the cosmological
perturbations  promote that the  comoving scale leaves the horizon
during inflation.
For models that have a standard reheating, this will correspond to
around 60 e-folds before the end of inflation. However, during
intermediate inflation the inflationary expansion never ends and
the model presents  the graceful exit problem. Equivalently, from
the point of view of the
 potential $V(\phi)$,  we observe that  this effective potential  does not present
a minimum,  wherewith  the usual mechanism introduced to achieve
inflation to an end becomes useless. As it  is well known,  the
standard reheating is described by the regime of oscillations of
the scalar field.  Since we do not know how the inflationary epoch
ends in intermediate law for the cold stages, one cannot draw any
further conclusions for this purpose, because the number of
$e$-folds to address the end of   inflation is unknown. A
methodology used in Refs.\cite{Teimoori:2017kob,MR} in order to solve
this problem consists in  introducing a determined time which
corresponds to unspecified reheating mechanism that triggered to
stop inflation. Here the number of $e$-folding at the moment of
horizon crossing is approximately 60 e-folds and the number of
e-folds to unspecified reheating mechanism  becomes  zero.

 In the following we will consider
the approximation made in
Refs.\cite{Barrow:1990vx,Barrow:1993zq,Barrow:2006dh} in order to
calculate the number of $e$-folds and the other
 cosmological parameters. Following Refs.\cite{Barrow:1990vx,Barrow:1993zq,Barrow:2006dh}
the number of $e$-folds $N$ between two different cosmic times
$t_{1}$ and $t_{2}$ or, equivalently   between two values of the
inflaton field $\phi_{1}$ and $\phi_{2}$, is given by
\begin{equation}
N=\int_{t_{1}}^{t_{2}}\,H\,dt=A\,\left(
t_{2}^{f}-t_{1}^{f}\right) =A\,k_1^{f}\left(
\phi_{2}^{\,f\mu_1}-\phi_{1}^{\,f\mu_1}\right)
 . \label{N1}%
\end{equation}
Here, we have used Eq.(\ref{sf1}).

In order to determine the beginning of inflationary phase, we find that
dimensionless slow-roll parameter
$\varepsilon_1=\varepsilon_1(\phi)$, is given by
\begin{equation}
\varepsilon_1=\left( {\frac{1-f}{Af}}\right) \,
k_1^{\,-f}\,\label{e1} \phi^{\,-f\mu_1}.
\end{equation}
In this sense, the condition for inflation takes place is given by
$\varepsilon_1<$1  (or equivalently $\ddot{a}>0$), then from
Eq.(\ref{e1}) the scalar field is such that $\phi>\left(
{\frac{1-f}{Af}}\right) ^{1 \over f\mu_1 }  k_1 ^{-1 \over \mu_1
}$ during inflation. Since inflation begins  at the earliest
possible scenario (see Fig.1), that is, when the slow-roll
parameter $\varepsilon_1(\phi=\phi_1)=\varepsilon_1(\phi_1)=1$ (or
equivalently  $\ddot{a}=0$), then the scalar field at the
beginning inflation $\phi_{1}$ results
\begin{equation}
 \phi_{1}=\left(
{\frac{1-f}{Af}}\right) ^{1 \over f\mu_1 }  k_1 ^{-1 \over \mu_1
}.\label{e11}
\end{equation}

 We note  that during the intermediate expansion the slow roll
parameter $\varepsilon_1$ in terms of the number of e-folds $N$
can be written as
\begin{equation}
\varepsilon_1=-\frac{\dot{H}}{H^2}=\frac{1-f}{1+f(N-1)}.\label{Nf}
\end{equation}
This suggests that the inflationary epoch begins at the earliest
possible stage when the number of $e$-folding is equal to $N=0$
(unlike Ref.\cite{Teimoori:2017kob}), in which the slow roll
parameter $\varepsilon_1\equiv 1$
\cite{Barrow:1990vx,Barrow:1993zq}.  In this context, in the
following we will evaluate the cosmological observables   in terms
of the number of $e$-folds $N$  which have took place since the
beginning of inflationary epoch, where the number of $e$-folding
at the moment of horizon crossing is approximately 50-70
$e$-folds. Also, note that for large $N$ such that $N\gg 1$, the
slow-roll parameter $\varepsilon_1\rightarrow 0$ and inflation
never ends in the cold models of intermediate expansion for the
case of a single field (inflaton).

 In relation to the initial value of  the Hubble parameter, we have that $H(t)=Af/t^{1-f}$
and the slow-roll parameter
$\varepsilon_1(t)=\frac{1-f}{Af}t^{-f}$. Thus,  we find that at
the earliest possible stage in which $\varepsilon_{1}(t=t_1)=1$,
the Hubble parameter at beginning of inflation  becomes
\begin{equation}
H(t=t_1)=H_1=\frac{(Af)^{1/f}}{(1-f)^{(1-f)/f}},\label{H11}
\end{equation}
where the initial value of $H_1\lesssim 1$
 (in
units of Planck mass)  from the classical description of the
universe. Here we note that the initial value of the Hubble
 rate $H_1$ depends on the values of the parameters $f$
and $A$.

In order to satisfy the condition ${\cal{A}}\gg 1$, we write the
parameter ${\cal{A}}$ in terms of the number of $e$-folds $N$ as
\begin{equation}
\mathcal{A}(N)=\mathcal{A}_0\,A\,f\, \left(1-f\right)^{2n-1 \over
2n+1} \left[{Af\over 1+f(N-1) }\right] ^{{2n-1 \over
f(2n+1)}+{1-f\over f}} [\phi(N)]^{2\nu \over 2n+1}\gg 1,\label{AAA}
\end{equation}
where $ \mathcal{A}_0={3\,n\,\gamma\over 2^{\,n-1}}\left[
{2^{\,n}\over 3\,n\,\gamma}\right] ^{2n-1 \over 2n+1} $ and the
scalar field $\phi(N)$ is defined as \be \phi(N)= \left({2n+1 +\nu
\over 2n} \right) ^{2n+1\over2n+1+\nu } \left({2^n \left(1-f
\right)\over 3n\gamma} \right) ^{1\over 2n+1+\nu}
\,\left[{1+f(N-1)\over Af }\right] ^{2n\over 2n+1+\nu}
\,\label{sf122}. \en
Here, we have used Eqs.(\ref{sf1}),
(\ref{N1}) and (\ref{e11}).

On the other hand, the scalar power spectrum
$\mathcal{P}_{\mathcal{S}}$ in terms of the scalar field reads as
\be \mathcal{P}_{\mathcal{S}}(\phi)={3\sqrt{3n}\over 32\pi^2
\sqrt{2} } {A^3f^3\over (1-f)} \, k_1^{\,-(2-3f)}\,
\phi^{\,-(2-3f)\mu_1} \,,\label{PR1} \en where we have used
Eqs.(\ref{P1}) and (\ref{sf1}), respectively. Now, from
Eqs.(\ref{N1}), (\ref{e11}) and (\ref{PR1}), we can write the
 scalar power spectrum as function of the number of
$e$-folds $N$ in the form \be
\mathcal{P}_{\mathcal{S}}(N)={3\sqrt{3n}\over 32\pi^2 \sqrt{2} }
{A^3f^3\over (1-f)}\left[{Af\over 1+f(N-1) } \right]^{2-3f\over f}
\,.\label{PRN1} \en Similarly, the scalar spectral index $n_s$ can
also be expressed in terms of the number $N$ as \be
n_s(N)=1-{2-3f\over 1+f(N-1) }.\label{ns1} \en Here, we noted that
the scalar spectral index given by Eq.(\ref{ns1}) coincides with
the obtained in the standard intermediate inflation
\cite{Barrow:2006dh}. Thus, for the special case in which $f=2/3$,
the scalar spectral index $n_s=1$ (Harrison-Zel'dovich spectrum).
In particular, assuming that the number of $e$-folds $N=60$ and
the spectral index $n_s=0.967$, we obtain that the value of the
parameter $f$ results $f=0.398\simeq0.4$.

From Eq.(\ref{r}), we find that the relationship between the
tensor-to-scalar ratio $r$ and the scalar spectral index $n_s$
results \be r(n_s)={64\sqrt{2}\over 3 \sqrt{3n} }
{(1-f)(1-n_s)\over (2-3f)
},\,\,\,\,\;\;\mbox{with}\,\,\,\,\;\;f\neq\frac{2}{3}.\label{rf}
\en Here, we note that the consistency relation $r(n_s)$ given by
(\ref{rf}) depends on the parameter $n$ through slope
$1/\sqrt{n}$, when compared to the results of $r(n_s)$ in the
standard intermediate model (recalled that $n\geq2/3$). Thus, this
dependence in the consistency relation ($\propto n^{-1/2}$) is
fundamental in order to the theoretical predictions enter inside
the allowed region of contour plot in the $r-n_s$ plane imposed by
 BICEP2/Keck-Array data, resurrecting the
intermediate inflation model.

\begin{figure}[th]
{{\vspace{-0
cm}{\includegraphics[width=2.3in,angle=0,clip=true]{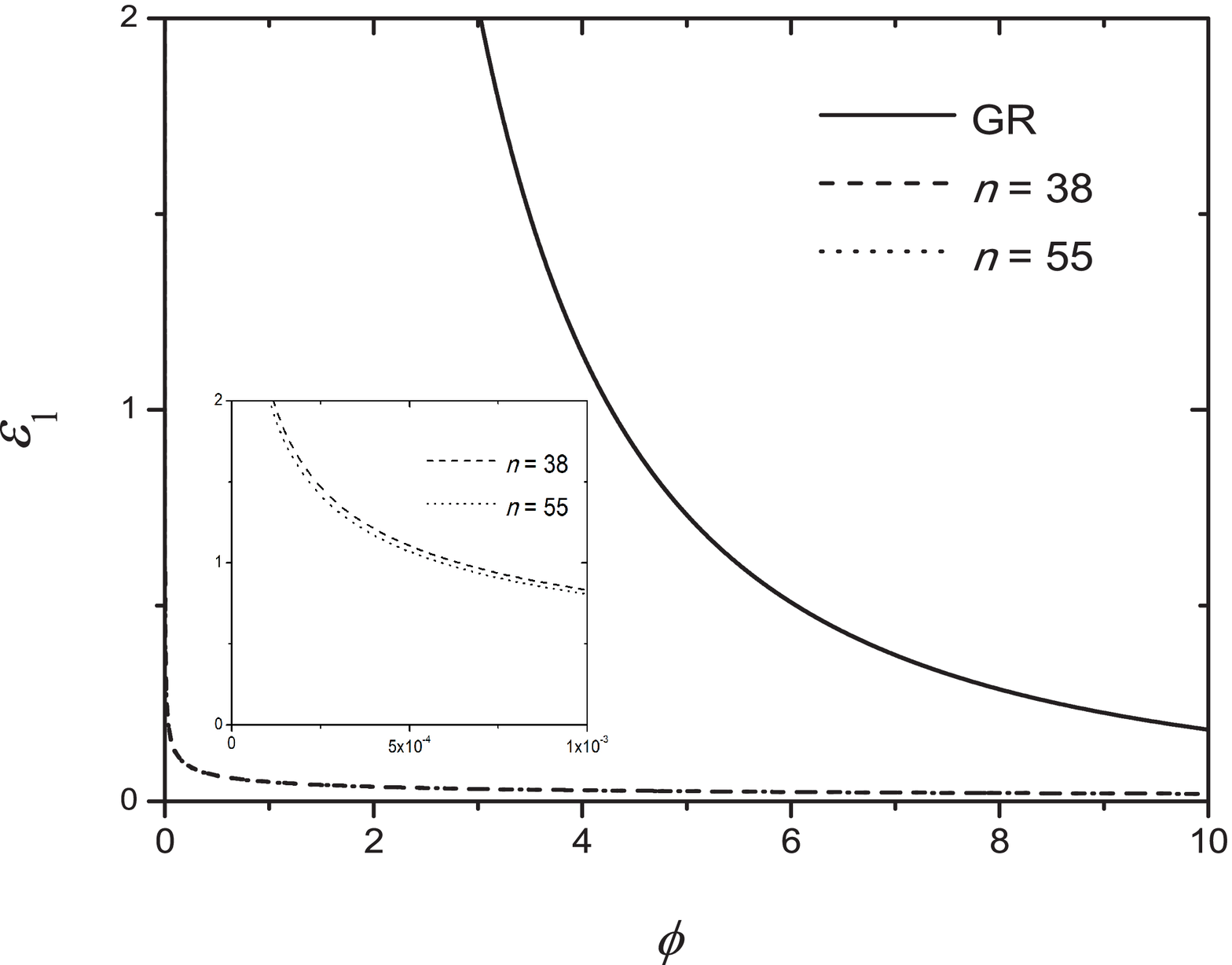}}}}
{{\vspace{-0
cm}\includegraphics[width=2.4in,angle=0,clip=true]{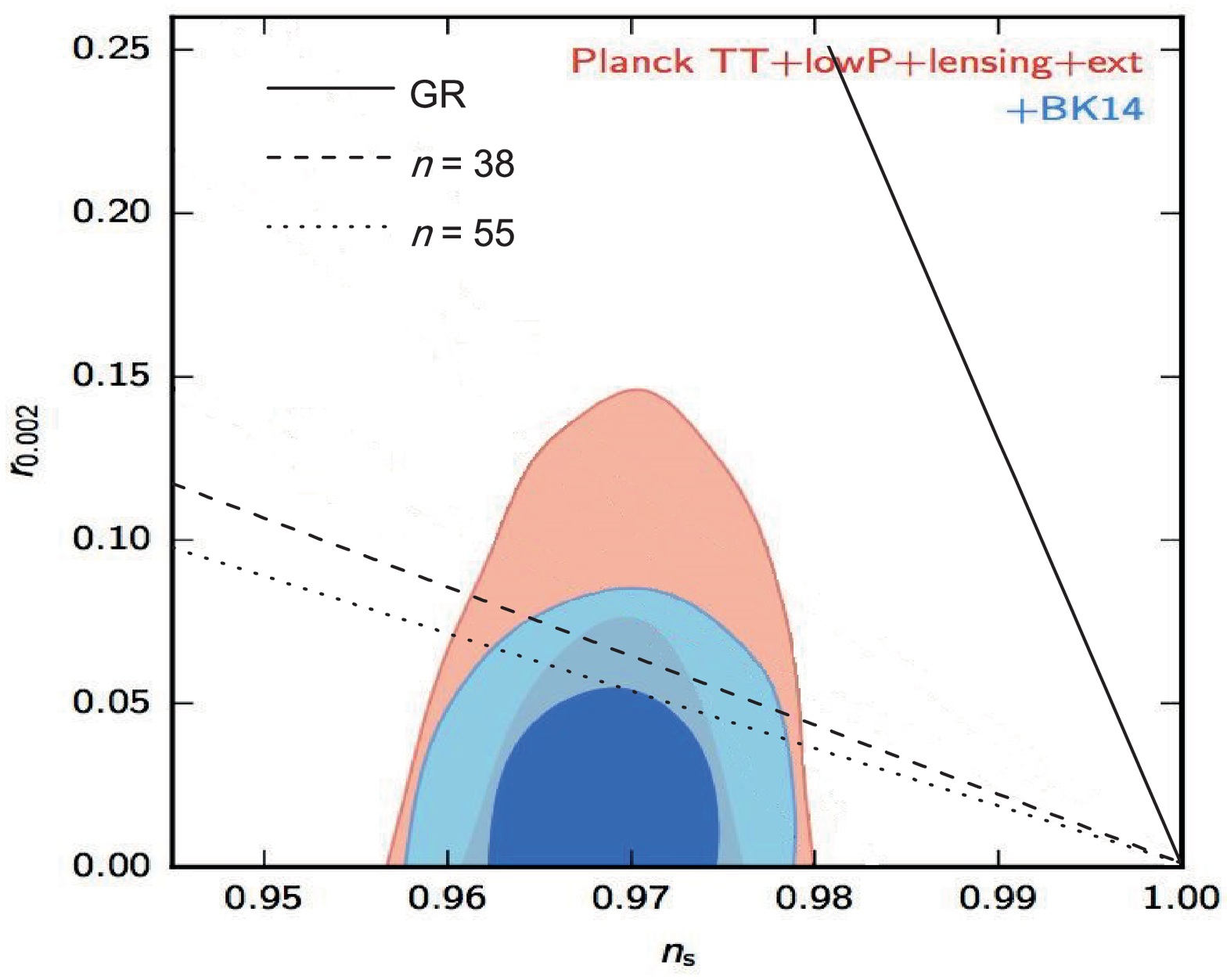}}}
{\vspace{0 cm}\caption{The dependence of the slow-roll parameter
$\varepsilon_1$ on the scalar field $\phi$ (left panel) and the
contour plot for the tensor-to-scalar ratio $r$ versus the scalar
spectral index $n_s$ (right panel). In left panel, we show that
the inflationary epoch never ends, since $\varepsilon_1\rightarrow
0$ for large $\phi$. In right panel we show  from BICEP2/Keck
Array Collaborations data, the two-dimensional marginalized
constraints (68$\%$ and 95$\%$ confidence
 levels) on the consistency relation $r(n_s)$ \cite{Array:2015xqh}.
 In both panels
and from left to right, dotted and
 dashed and lines correspond to the cases where the power $n$ takes the values $n=55$ and $n=38$, respectively. Finally, the solid line corresponds
 to the standard intermediate model.
 In these plots we have used $f=0.4$.
 \label{fig1}}}
\end{figure}

From BICEP2/Keck-Array results  data that the ratio $r<0.07$, we
find a lower bound for the power $n$ given by
$n>61912(1-f)^2(1-n_s)^2/(2-3f)^2$. In particular, for the values
$f=0.4$ and $n_s=0.967$, the lower limit for $n$ yields $n>38$.
Also, we note that from Eq.(\ref{PRN1}), we can find a constraint
for the parameter $A$ of the intermediate model for given values
of $f$ and the power $n$, when the number of $e$-folds $N$ and the
amplitude of the scalar power spectrum $\mathcal{P}_{\mathcal{S}}$
are also given. Thus, in particular for the values
$\mathcal{P}_{\mathcal{S}}=2.2\times 10^{-9}$, $N=60$ and $f=0.4$,
 we found that for $n=38$, $A$ becomes $A=0.26$, while
for the case $n=55$, we found that $A=0.25$.  In relation to the initial value of the Hubble parameter
$H_1$,
 we find by considering  Eq.(\ref{H11}) that for the value $n=38$, (where $A=0.26$ and $f=0.4$)  corresponds
to $H_1=7.5 \times 10^{-3}$ (in units of Planck mass) and for the case in which $n=55$
 (in which $A=0.25$ and $f=0.4$) we have $H_1=6.8\times 10^{-3}$.
In addition, from the
condition ${\cal{A}}\gg 1$
 given by Eq.(\ref{AAA}), we are able to find a lower bound for the parameter $\gamma$, for different values of the parameter $\nu$,
 when  the number of $e$-folding $N$, $f$ and $n$ are given.
Here, we mention that the parameter ${\cal{A}}$ satisfies the
condition ${\cal{A}}=3ng(\phi)X^{n-1}H\dot{\phi}\gg 1$ as
$g(\phi)\gg (3nX^{n-1}H\dot{\phi})^{-1}$.   In order to give
an estimation for the coupling parameter $g$, we have that
typically after of started the inflationary epoch, the Hubble rate
 $H\sim 10^{-5}$ and $\dot{\phi}\sim 10^{-5}$, thus we find that
the coupling $g$ has a lower bound given by $g(\phi)\gg 10^{400}$
for $n\sim40$. This suggests that the coupling $g(\phi)$ must have
a very large value as lower bound (googol$^4$).
  In particular for the $N=60$, $f=0.4$ and $n=38$, and since that $g(\phi)=\gamma\phi^{\nu}$,  we find that for the case $\nu=1$,
 the lower limit is found to be $\gamma\gg 8\times 10^{403}$, while for $\nu=0$ (or equivalently $g(\phi)=$ const.)
 we have that $\gamma\gg 10^{404}$. Finally, for the case $\nu=-1$ (or $g(\phi)\propto \phi^{-1}$),  we found that $\gamma\gg 10^{405}$.

In Fig.\ref{fig1}, the left panel shows the evolution of the
slow-roll parameter $\varepsilon_1$ in terms of the scalar field
$\phi$, while the right panel shows the contour plot for the
consistency relation $r(n_s)$. In both panels, we consider the
cases where the power $n$ has two different values in addition to
the standard intermediate model. Here we have used the value
$f=0.4$. In order to write down values for the slow-roll parameter
$\varepsilon_1(N)$ and the ratio $r=r(n_s)$, we have used
Eqs.(\ref{N1}), (\ref{e1}) and (\ref{rf}), respectively. From left
panel we show that the inflationary epoch never ends in the
G-intermediate model (in the same form  as it occurs in standard
intermediate model), since during inflation the slow-roll
parameter $\varepsilon_1$ always is $\varepsilon_1<1$ and tends to
$\varepsilon_1\rightarrow 0$ for large $\phi$, see Fig.\ref{fig1}
(left panel). In this sense, we consider that inflationary stage
begins at the earliest possible scenario when
$\varepsilon_1(\phi=\phi_1)=1$, where $\phi_1$ is given by
Eq.(\ref{e1}). Here, we have shown that the authors of
Ref.\cite{Teimoori:2017kob}  committed a mistake when they
computed the time at which inflation ends  in the intermediate
G-model, since inflation never ends. As it can visualized  from
right panel of Fig.\ref{fig1}, for values of the power $n$
satisfying $n>38$, the model is well supported by the data. Also,
we noted that when $n\gg 1$, then the tensor-to-scalar ratio
$r\sim 0.$

 On the other hand, the predictions for the intermediate model
regarding primordial NG, for the particular case $n=38$, we find
that the values of $f_{NL}$ in the cases;  equilateral,
orthogonal, and enfolded configurations become
$f_{NL}^{\textup{equil}}=-8.17$, $f_{NL}^{\textup{ortho}}=-5.48$,
and $f_{NL}^{\textup{enfold}}=-1.34$, respectively. Finally, for
$n=55$, we have that $f_{NL}^{\textup{equil}}=-11.95$,
$f_{NL}^{\textup{ortho}}=-8.03$, and
$f_{NL}^{\textup{enfold}}=-1.96$, respectively. Here, we check
that the primordial NG $\mid f_{NL}\mid\gtrsim 1$. In this sense,
these values are within the current observational bounds set by
Planck.

\section{Logamediate G-inflation
}\label{Glog}

Now, we consider the situation in which the scale factor evolves
according to logamediate inflation, given by Eq.(\ref{loga}).
Here, the Hubble rate $H(t)$ becomes $H(t)=B\lambda{(\ln
t)^\lambda\over t}$, and from Eq.(\ref{A1}), we find that the
scalar field $\phi(t)$ results
 \be \phi(t)= \left({2n+1 +\nu \over
2n} \right) ^{2n+1\over2n+1+\nu } \left({2^n \over 3n\gamma}
\right) ^{1\over 2n+1+\nu} \,\left( t^{2n\over 2n+1}-1\right)
^{2n+1\over2n+1+\nu } \,.\label{sf21} \en By assuming the slow-roll equation (\ref{frwa2}), we have that the effective
potential in terms of the scalar field is given by
\begin{equation}
V(\phi)=V_0\, (1+k_2\,\phi^{\mu_2})^{-{2n+1\over n} }
[\ln(1+ k_2\, \phi^{\,\mu_2})]^{2\lambda},
\end{equation}
where the constants $V_0$, $k_2$ and $\mu_2$ are defined as
$$
V_0=3(B\lambda)^2,\;\;\;k_2=\left({2n \over 2n+1 +\nu } \right) \left({3n\gamma\over 2^n }
\right) ^{1\over
2n+1},\,\,\,\,\mbox{and}\,\;\;\;\;\;\mu_2={2n+1+\nu \over 2n+1},
$$
respectively. For the logamediate expansion in the context of
G-inflation, the number of $e$-folds $N$ between two different
values of the scalar field $\phi_1$ and $\phi_2$  is written
as
\begin{equation} N=B\left[(\ln t_{2})^{\lambda}-(\ln
t_{1})^{\lambda}\right] =B\left( {2n \over 2n+1 }\right)
^{\lambda}\left(  [\ln(1+ k_2\,
\phi_2^{\,\mu_2})]^{\lambda}-[\ln(1+ k_2\,
\phi_1^{\,\mu_2})]^{\lambda}]\right). \label{NL1}
\end{equation}
Here, we have used Eq.(\ref{sf21}).

As before, we write ${\cal{A}}(N)$ in order to satisfy the
condition ${\cal{A}}\gg 1$. Thus, we have that ${\cal{A}}(N)$
becomes
\begin{equation}
\mathcal{A}(N)=\mathcal{A}_0\,B \lambda [\Xi(N)]^{\lambda-1}
e^{-{4n \over 2n+1}\Xi(N)} [\phi(N)]^{2\nu \over 2n+1}\gg 1,\label{NN2}
\end{equation}
where the field $\phi(N)$ and the function $\Xi(N)$ are defined as
$$ \phi(N)= \left({2n+1 +\nu \over 2n} \right) ^{2n+1\over2n+1+\nu
} \left({2^n \over 3n\gamma} \right) ^{1\over 2n+1+\nu} \,\left(
e^{{2n\over 2n+1}\Xi(N)}-1\right) ^{2n+1\over2n+1+\nu },
$$
and
$$
 \Xi (N)=\left[{N\over B }+\left( {1\over B \lambda}\right) ^{\lambda\over \lambda-1}\right]^{1 \over
 \lambda},
$$
respectively.

For the dimensionless slow-roll parameter $\varepsilon_1$ in the
logamediate G-inflation, we have that
$$
 \varepsilon_1=\left(
{\frac{1}{B\lambda}}\right)\left({2n \over 2n+1  } \right)
^{\lambda-1}
 \,[\ln(1+ k_2\, \phi^{\,\mu_2})]^{1-\lambda},
$$
and in order to get an inflationary scenario ($\varepsilon_1<$1),
we have that the scalar field
 $\phi>k_2 ^{-1 \over \mu_2 }
\left( \exp\left[ {2n \over 2n+1 }(B\lambda)^{{-1\over
\lambda-1}}\right] -1\right) ^{1 \over f\mu_2} $. As before, if
 the  inflationary stage begins at the
earliest possible epoch, where the slow-roll parameter
$\varepsilon_1(\phi=\phi_1)=1$,
 then we obtain that the
 field $\phi_{1}$ is given by
\begin{equation}
 \phi_{1}= k_2 ^{-1 \over \mu_2 }
\left( \exp\left[ {2n \over 2n+1 }(B\lambda)^{{-1\over
\lambda-1}}\right] -1\right) ^{1 \over f\mu_2}.\label{ee1}
\end{equation}

 For this expansion,  the Hubble rate $H(t)$ is given by
$H=B\lambda(\ln t)^{\lambda-1}/t$ and the slow roll parameter
$\varepsilon_1(t)=(B\lambda)^{-1}\ln t^{1-\lambda}$, thus we find
that at the earliest possible stage in which
$\varepsilon_{1}(t=t_1)=1$, the Hubble parameter at beginning of
inflation  becomes
$H(t=t_1)=H_1=\exp[-(1/B\lambda)^{1/(\lambda-1)}]$, and this
initial rate depends exclusively  on the associated  parameters
$B$ and $\lambda$ of the scale factor.

On the other hand, as before we find that the scalar power
spectrum $\mathcal{P}_{\mathcal{S}}$ as function of the number of
$e$-folds reads as \be
\mathcal{P}_{\mathcal{S}}(N)={3\sqrt{3n}\over 32\pi^2 \sqrt{2} }
B^3\lambda^3 e^{-2\left[{N\over B }+\left( {1\over B
\lambda}\right) ^{\lambda\over \lambda-1}\right]^{1\over \lambda}}
\left[{N\over B }+\left( {1\over B \lambda}\right) ^{\lambda\over
\lambda-1}\right]^{3(\lambda-1)\over \lambda} \,.\label{PRN2} \en
Here, we have considered Eqs.(\ref{P1}), (\ref{ee1}) and
(\ref{NL1}).

\begin{figure}[th]
\includegraphics[width=3.5in,angle=0,clip=true]{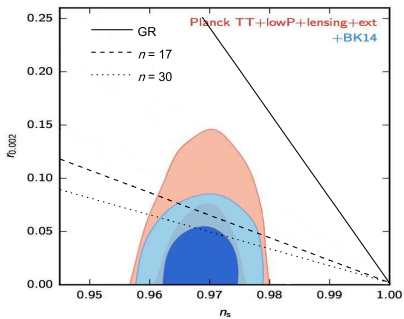}
{\vspace{-1 cm}\caption{ The contour plot for the tensor-to-scalar
ratio $r$ versus the scalar spectral index $n_s$, for the
logamediate expansion in the context of G-inflation. In this plot
and from left to right,  dotted and dashed lines correspond to the
cases when the power $n$ has the values $n=30$ and $n=17$,
respectively. Solid line corresponds to the standard logamediate
inflation model. Here, from BICEP2/Keck Array Collaboration, the
two-
 dimensional marginalized constraints (68$\%$ and 95$\%$ confidence
 levels) on the consistency relation $r(n_s)$ \cite{Array:2015xqh} are shown.
 \label{fig2}}}
\end{figure}

Now, from Eqs.(\ref{ns}), (\ref{NL1}) and (\ref{ee1}), we find
that the scalar spectral index $n_s$ is related to the number of
$e$-folds $N$ through the following expression \be
n_s(N)=1-{2\over B\lambda }\left[{N\over B }+\left( {1\over B
\lambda}\right) ^{\lambda\over
\lambda-1}\right]^{-(\lambda-1)\over \lambda}.\label{ni} \en Note
that this expression for the scalar spectral index coincides with
the obtained from logamediate inflation in GR
\cite{Barrow:2007zr}.

In a similar fashion as we did before, we find that the
consistency relation $r=r(n_s)$ is given by \be
r(n_s)={32\sqrt{2}\over 3 \sqrt{3n} } (1-n_s).\label{rrr3}
 \en
As in the previous case of intermediate G-inflation, we noted that
the  relation $r=r(n_s)$ given by (\ref{rrr3}) strongly depends on
the power $n$,  when we make the comparison with the results of
$r(n_s)$ in the standard logamediate model in the framework of GR.
In this sense, the  dependence on the power $n$ is crucial in
order for the theoretical predictions of the model  to enter
 in the allowed regions of the contour plot in the
$r-n_s$ plane.  We also note that, for  large values of the power
$n$ such that $n\gg 1$, the tensor-to-scalar ratio $r$ tends to
zero. From BICEP2/Keck-Array data, we have that $r<0.07$,  then we
find a lower bound for the power $n$, given by $n>15480(1-n_s)^2$.
In particular, considering that the scalar spectral index takes
the value $n_s=0.967$, the lower limit for the power $n$ yields
$n>17$.

 Also, from Eqs.(\ref{PRN2}) and (\ref{ni}),
we  may find a constraint for the parameters $B$ and $\lambda$,
 appearing in the logamediate model, when the
power $n$, the number of $e$-folds $N$,  the power spectrum
$\mathcal{P}_{\mathcal{S}}$ as well as $n_s$ are given.
 Particularly, for $N=60$ and considering the
observational values $\mathcal{P}_{\mathcal{S}}=2.2\times 10^{-9}$
and $n_s=0.967$, we found the values $B=6.2\times 10^{-16}$ and
$\lambda=14.6$ when the power $n$ is fixed to be $n=17$. On the
other hand, for the case when $n=30$, we obtain the values
$B=3.9\times10^{-16}$ and $\lambda=14.7$.  In order to
determine the initial value of the  Hubble rate $H_1$, we have
that for the case $n=17$, where $B=6.2\times 10^{-6}$ and
$\lambda=14.6$ , we find that $H_1=2.1\times10^{-5}$ (in units of
Planck mass) and for the case in which $n=30$ corresponds to
$H_1=1.8\times 10^{-5}$.

Besides, considering the
condition ${\cal{A}}\gg 1$,
 given by Eq.(\ref{NN2}), we find a lower bound for the parameter $\gamma$ as in the case of intermediate
 inflation, by assuming different
 values of the parameter $\nu$,
 when  the number of $e$-folding $N$, $\lambda$ and  the power $n$ are given.
  In particular, by fixing $N=60$, $\lambda=14.6$, $n=17$, for $\nu=1$ the lower
  limit on $\gamma$ is found to be $\gamma\gg 5\times10^{181}$, while for $\nu=0$ (or equivalently $g(\phi)=$ constant)
 we have that $\gamma\gg 6\times10^{182}$. Finally, for $\nu=-1$ (or $g(\phi)\propto \phi^{-1}$), the lower limits yields $\gamma\gg 6\times10^{183}$.

In Fig.\ref{fig2}, we show the contour plot together with the
consistency relation $r(n_s)$. In this panel we consider two
different values of the parameter $n$ in the G-logamediate model
and also we show the standard logamediate model. Here we have used
the corresponding pair of values ($B$,$\lambda$) for a given value
of the power $n$. Note that for values  of the power $n$
satisfying $n>17$, the model is well supported by current data, as
it can be seen from Fig.\ref{fig2}.
 Moreover, as in the intermediate model,
for large values of the power $n\gg 1$, the tensor-to-scalar ratio
$r\sim 0.$  Also, by considering the lower bound on $n$ for
this model, the predicted values for $f_{NL}$ in the equilateral,
orthogonal, and enfolded configurations become
$f_{NL}^{\textup{equil}}|_{n=17}=-3.50$,
$f_{NL}^{\textup{ortho}}|_{n=17}=-2.34$, and
$f_{NL}^{\textup{enfold}}|_{n=17}=-0.58$,
 respectively. We also mention that for values of $n>29$, the
 primordial NG $|f_{NL}|\gtrsim 1$. Thus, for values of $n\geqslant 17$, we find that
 parameter $|f_{NL}|$
 is in  well agreement with current observational data.

\section{Exponential G-inflation }\label{Gexp}
Now, we study the case in which the Hubble rate is given by
 $H(t)=\alpha\,e^{-\beta \,t}$, where the parameters $\alpha$ and $\beta$ are
 positive constants. From Eq.(\ref{A1}) we obtain that the scalar
field $\phi$ as function of the cosmic time becomes \be \phi(t)=
\left({2n+1 +\nu \over 2n+1} \right) ^{2n+1\over2n+1+\nu }
\left({2^n \beta \over 3n\gamma} \right) ^{1\over 2n+1+\nu}
\,t^{2n+1\over2n+1+\nu } \,.\label{sf3} \en From the Friedmann
equation (\ref{frwa2}), we find that the effective potential in
terms of the scalar field can be written as
$$
V(\phi)=V_1\,e^{-2\beta
\,k_3\,\phi^{\mu_3}},\,\,\;\;\,\,\mbox{with}\,\,\,\,\,\;\;V_1=3\alpha^3,
$$
where the constants $k_3$ and $\mu_3$  are defined as
%
$$
k_3=\left({2n+1 \over 2n+1 +\nu } \right)
\left({3n\gamma\over 2^n \beta} \right) ^{1\over
2n+1},\,\;\;\;\mbox{and}\,\,\,\,\,\mu_3={2n+1+\nu \over 2n+1}=\mu_2,
$$
respectively.

For this Hubble rate, the number of $e$-folds $N$ between two different
values of the scalar field $\phi_1$ and $\phi_2$ results
\begin{equation}
N=\int_{t_1}^{t_{2}}\,H\,dt={\alpha \over \beta}\,\left[e^{-\beta \,t_1}-e^{-\beta \,t_2}\right]
={\alpha \over \beta}\,\left[e^{-\beta \,k_3\,\phi^{\mu_3}_1}-e^{-\beta \,k_3\,\phi^{\mu_3}_2}\right]. \label{NE1}
\end{equation}

$\cal{A}$ as function of the number of e-folding $N$ can be written
as
\be
\mathcal{A}(N)=\mathcal{A}_0\,
\beta^{{4n \over 2n+1}}
(N+1)
[\phi(N)]^{2\nu \over 2n+1}\gg 1,\label{RR3}
\en
where the scalar field $\phi(N)$ reads as
$$
\phi(N)= \left({2n+1 +\nu \over 2n+1} \right) ^{2n+1\over2n+1+\nu }
\left({2^n \beta \over 3n\gamma} \right)
^{1\over 2n+1+\nu}
\,\beta^{-{2n+1\over2n+1+\nu } }
\left( \ln\left[{\alpha \over \beta(N+1) } \right] \right) ^{{2n+1\over2n+1+\nu }
}.
$$

 Unlike the intermediate and logamediate inflation models, this Hubble rate addresses the end of the accelerated expansion. In
this sense, considering that inflation ends when $\varepsilon_1=1$, where the
slow-roll parameter $\varepsilon_1$ is given by
$$
\varepsilon_1=
{\frac{\beta}{\alpha}}
\,\exp(\beta k_3\, \phi^{\,\mu_3}),
$$
we have that the scalar field at the end of inflation, given by $\varepsilon_1(\phi=\phi_2)=1$, becomes
$$ \phi_{2}= (\beta k_3 )^{-1 \over \mu_3 }
\left[ \ln( \alpha / \beta ) \right]
^{1 \over \mu_3}.
$$
 Since during the exponential  expansion, the inflationary
scenario ends, then   the Hubble rate $H(t)$ is given by
$H=\alpha\exp[-\beta t]$ and the slow-roll parameter
$\varepsilon_1(t)=\beta\exp[\beta t]/\alpha$. Thus, we find that
at the end of inflation in which $\varepsilon_{1}(t=t_2)=1$, the
Hubble parameter at this time  becomes $H(t=t_2)=H_2=\beta$.

Also, from
 the condition for
inflation to occur
in which
$\varepsilon_1<$1, then the scalar field becomes $\phi< (\beta k_3 )^{-1 \over \mu_3 }
\left[ \ln( \alpha / \beta ) \right]
^{1 \over \mu_3}  $.

As before, we can express the the amplitude of the scalar power
spectrum $\mathcal{P}_{\mathcal{S}}$ in terms of the number of
e-folding $N$ as \be \mathcal{P}_{\mathcal{S}}(N)={3\sqrt{3n}\over
32\pi^2 \sqrt{2} }\beta(N+1)^3 \,,\label{PRN3} \en and the scalar
spectral index $n_s(N)$ results $ n_s(N)=1-{3\over N+1 } $. Also,
we find that the consistency relation $r=r(n_s)$ in this scenario
can be written as \be r(n_s)={64\sqrt{2}\over 9 \sqrt{3n} }
(1-n_s).\label{er1} \en As in the previous models of intermediate
and logamediate, we observed that the consistency relation
$r=r(n_s)$ given by (\ref{er1}) also strongly depends on the power
$n$. As before, the introduction of  the power $n$ in the model is
fundamental in order to  the theoretical predictions of this model
enter in the allowed region of the contour plot in the $r-n_s$
plane from \cite{Array:2015xqh}. Assuming the BICEP2/Keck-Array,
for which $r<0.07$, we obtain a lower bound for the power $n$,
given by $n>6880(1-n_s)^2$. In particular assuming that the scalar
spectral index $n_s$ is given by $n_s=0.967$, we find that the
lower bound for the power $n$ corresponds to $n>7$.

  In addition,  from the the amplitude of the scalar power spectrum given by
eq.(\ref{PRN3}), we can find a constraint for the parameter
$\beta$, appearing in the Hubble rate, for several values of $n$
when the number of $e$-folds $N$ and the observational value of
the power spectrum $\mathcal{P}_{\mathcal{S}}$ are given. Thus,
particularly for the values
$\mathcal{P}_{\mathcal{S}}=2.2\times10^{-9}$ and $N=60$,
 for the case when the power $n$ takes the value
$n=8$,  we found the value $\beta=2.9\times10^{-13}$. As in the
previous models, we can find a lower bound
 for the parameter $\gamma$
from  the condition ${\cal{A}}\gg 1$
 given by Eq.(\ref{RR3}).  In particular,  for the values
$N=60$, $\beta=2.9\times10^{-13}$, $\alpha=10^{-3}$ and $n=8$, we obtain that for the case in which $\nu=1$
 ($g(\phi)\propto \phi$), the lower bound is $\gamma\gg3\times 10^{181}$,
  while for $\nu=0$ (or $g(\phi)=$ constant) we have that $\gamma\gg 6\times 10^{183}$.
  Finally, for the specific case in which $\nu=-1$ (or $g(\phi)\propto \phi^{-1}$), we obtain that
   $\gamma\gg 10^{186}$. As in the previous models, from the two-dimensional marginalized constraints on the $r-n_s$ plane, this model becomes
well supported by the Planck data when the power $n$ satisfies
$n>7$ (figure not shown) and then the model works.  We also
mentioned that as the Hubble parameter at the end of inflation, is
given by $H_2=\beta$, then this rate at that time becomes
$H_2<\frac{0.1\pi^2\mathcal{P}_{\mathcal{S}}}{(N+1)^3(1-n_s)^2}$.
Here we have used eq.(\ref{PRN3}) and the fact that
$n>6880(1-n_s)^2$. In particular, for the values
${\mathcal{P}}_{\mathcal{S}}=2.2\times 10^{-9}$, $N=60$ and
$n_s=0.967$, we have that the lower bound for the Hubble parameter
at the end of the inflationary epoch results $H_2<3\times
10^{-13}$ (in units of Planck mass).   In relation to the
primordial NG, we  obtain that for the lower bound of $n$, we have
that
 $f_{NL}^{\textup{equil}}|_{n=7}=-1.27$, $f_{NL}^{\textup{ortho}}|_{n=7}=-0.85$,
 and $f_{NL}^{\textup{enfold}}|_{n=7}=-0.21$, respectively. Thus,
 for values of the power $n>7$, the non-lineal parameter $f_{NL}$ is well corroborated by
 Planck data.

\section{Conclusions \label{conclu}}

In this paper we have investigated the intermediate, logamediate
and exponential inflation in the framework of a Galilean action
with  a coupling of the form $G(\phi,X)\propto \phi^\nu\,X^n$. For
a flat FRW universe, we have found solutions to the background and
perturbative dynamics for each of these expansion laws under the
slow-roll approximation. In particular, we have obtained explicit
expressions for the corresponding  scalar field, effective
potential, number of $e$-folding  as well as for the scalar power
spectrum, scalar spectral index and tensor-to-scalar ratio. In
order to bring about some analytical solutions, we have considered
that the Galileon effect dominates over the standard inflation, in
which the parameter ${\cal{A}}=3H\dot{\phi}G_X$ satisfies the
condition ${\cal{A}}\gg 1$. In this context, we have found
analytic expressions for the constraints on the $r-n_s$ plane, and
for all these G-inflation models we have obtained that the
consistency relation $r=r(n_s)$ depends on the power $n$ which is
crucial in order to the corresponding theoretical predictions
enter on the two-dimensional marginalized constraints imposed by
current BICEP2/Keck-Array data.  In this sense, we have
established that the inflationary models of intermediate,
logamediate and exponential in the framework of G-inflation are
well supported by the data, as could be seen from Figs.(1) and
(2). In particular for the intermediate G-inflation, from the
$r-n_s$ plane, we have found a lower bound for  the power $n$,
given by $n>38$. For the logamediate model we have obtained that
$n>17$ and finally, for the exponential model we have got $n>7$ as
lower limit. Also, we have found that for values of $n\gg 1$, the
tensor-to-scalar ratio $r\rightarrow 0$. Also, from the
 amplitude of the scalar power spectrum
$\mathcal{P}_{\mathcal{S}}(N)$ and the scalar spectral index
$n_s(N)$ as function of the number of $e$-folds, we have found
constraints on the several parameters appearing in our models.
Besides, considering that the Galileon effect dominates over GR
given by the condition ${\cal{A}}\gg 1$,  we have found a very
large value as a lower limit for the parameter $\gamma$. The
reason for this is due that typically $H\sim\dot{\phi}\sim
10^{-5}\ll 1$, then from the condition ${\cal{A}}\gg 1$ suggesting
$g(\phi)\gg (3nX^{n-1}H\dot{\phi})^{-1}$, thus we have found that
$g(\phi)\gg (3nX^{n-1}H\dot{\phi})^{-1}\sim
\mathcal{O}(10^{400})$, e.g. for $n=40$, and for $n\sim 10$ we
have got $g(\phi)\gg (3nX^{n-1}H\dot{\phi})^{-1}\sim
\mathcal{O}(10^{100})$ (googol).
In relation to the primordial
NG, we have found that  for limit in which the Galilean dominated
regime i.e., ${\cal{A}}\gg 1$, the non-linear parameter $\mid
f_{NL}\mid\propto n$ and it is within the current observational
bounds imposed by  Planck data.

In this work, we have determined that the intermediate,
logamediate and exponential models in the context of G-inflation,
are less restricted than those in the framework of standard GR,
due to the modification in the action by the Galilean term
$G(\phi,X)\square\phi\propto\phi^{\nu}\,X^n\square\phi$.

Finally, in this paper we have not addressed a mechanism to bring
intermediate and logamediate G-inflation to  an end and therefore
to a study the mechanism of reheating, see
Refs.\cite{Herrera:2016sov,delCampo:2007iw}. Also, we have
not guided our investigation
 on the  non-canonical
K-inflation terms in order to discern its importance in relation
to the cubic Galileon term for these expansions. We hope to return
to address these points for these models of G-inflation in the
near future.

\begin{acknowledgments}
R.H. was supported by Proyecto VRIEA-PUCV N$_{0}$ 039.309/2018.
N.V. acknowledges support from the Fondecyt de Iniciaci\'on project N$^o$ 11170162.
\end{acknowledgments}


\end{document}